\documentclass[aps,prd,twocolumn,groupedaddress, showpacs]{revtex4} 

\usepackage{graphicx}

\begin{document}

\title{Cosmological constraints on the curvaton web parameters}

\author{Edgar Bugaev}
\email[e-mail: ]{bugaev@pcbai10.inr.ruhep.ru}

\author{Peter Klimai}
\email[e-mail: ]{pklimai@gmail.com}
\affiliation{Institute for Nuclear Research, Russian Academy of
Sciences, 60th October Anniversary Prospect 7a, 117312 Moscow, Russia}




\begin{abstract}
We consider the mixed inflaton-curvaton scenario in which quantum fluctuations of
the curvaton field during inflation lead to a relatively large curvature
perturbation spectrum at small scales.
We use the model of chaotic inflation with quadratic potential including
supergravity corrections leading to a large positive tilt in the power spectrum
of the curvaton field. The model is characterized by the strongly
inhomogeneous curvaton field in the Universe and large non-Gaussianity
of curvature perturbations at small scales.
We obtained the constraints on the model parameters considering the
process of primordial black hole (PBH) production in radiation era.
\end{abstract}


\pacs{98.80.-k, 04.70.-s  \hfill arXiv:1212.6529 [astro-ph.CO]}

\maketitle

\section{Introduction}
\label{sec-intro}

Curvaton mechanism which has been suggested $\sim$ 15 years ago
\cite{Mollerach:1989hu, Linde:1996gt, Lyth:2001nq, Moroi:2001ct, Lyth:2002my}
now is the object of intense study.
It is assumed, in the standard implementation of the curvaton model, that
not the inflaton field perturbations are responsible for the primordial density
fluctuations and for the cosmic microwave background fluctuations, but
instead the (isocurvature) perturbations of the curvaton field $\sigma$.
It is assumed that this curvaton field is subdominant during inflation
but in post-inflationary epoch when Hubble constant becomes small,
$H\sim m$ (where $m$ is the curvaton mass), curvaton starts oscillating
in its potential and behaves as nonrelativistic matter. The energy density
of the curvaton decreases as $\sim a^{-3}$ ($a$ is the scale factor)
whereas the energy density of radiation produced by the inflaton decay
decreases as $a^{-4}$. As a result the curvaton energy density grows
relative to radiation energy density until the curvaton contribution
becomes significant. If it happens before the curvaton decay one can
say that curvaton mechanism is ``effective'', in a sense that just the
curvaton (rather than inflaton) field perturbations during inflation
determine the resulting (adiabatic) curvature perturbations at cosmological
scales.

In scenarios with the ``effective'' curvaton there is the strong constraint
on a value of the curvaton mass: it must be much smaller than the Hubble
constant during inflation, $H_i$, otherwise the primordial density
perturbations have too large spectral tilt. Moreover, if the ratio
$m^2/H_i^2$ is not small, the coherent length of the curvaton field
(i.e., the characteristic size of the region inside of which the field
is approximately homogeneous) is also too small and, in particular,
smaller than the current horizon size. In the latter case, the primordial
perturbation spectrum is strongly non-Gaussian, in contradiction with
observations.

The condition $m^2/H_i^2 \ll 1$ is too restrictive and prohibits
an use, for a description of the curvaton, particle physics models
predicting large ratios $m^2/H_i^2$ at inflation (e.g., some
variants of supersymmetric theories).
In this connection it is reasonable to consider also
the mixed curvaton-inflaton scenarios \cite{Langlois:2004nn, Ferrer:2004nv}
in which the curvaton perturbations are additional to the usual perturbations
produced by the inflaton. Combining two contributions, one can
obtain the primordial perturbation spectrum which is in agreement with data
at cosmological scales. At the same time, the prediction for smaller
scales may be quite unusual: the spectrum can be, e.g., very blue
(i.e., the spectral tilt is large and positive) and, besides, the perturbations
can be strongly non-Gaussian. In particular, large value of the tilt
arises due to non-renormalizable and supergravity corrections to the Lagrangian
of some supersymmetric theories inducing mass terms of the order $H^2$
\cite{Dine:1983ys, Coughlan:1984yk, Goncharov:1984qm, Bertolami:1987xb, Dine:1995kz}.

In most curvaton scenarios it is assumed that the curvaton field in the
Universe is highly homogeneous and, as a result, the non-Gaussianity is
relatively small. According to the alternative hypothesis, after the long inflationary expansion,
the average value of the curvaton field is close to zero, and the local value of the
field has a Gaussian probability distribution, variance of which is
given by the formula \cite{BD, St1982}
\begin{equation}
\label{sigma2}
\langle \sigma^2({\bf x}) \rangle = \frac{3 H_i^4}{8\pi^2 m_*^2}=
\left(\frac{H_i}{2\pi} \right)^2 \frac{1}{t_\sigma} .
\end{equation}
Here, $m_*$ is the effective curvaton mass which differs from the true
curvaton mass $m$ \cite{Lyth:2004nx}. The corresponding coherent
length is
\begin{equation}
\label{ellc}
\ell_c \sim \frac{1}{H_i} \exp\left({\frac{3 H_i^2}{2 m_*^2}}\right) =
\frac{1}{H_i} \exp\left({\frac{1}{t_\sigma}}\right).
\end{equation}
In Eqs. (\ref{sigma2}) and (\ref{ellc}), $t_\sigma$ is the spectral tilt of the
perturbation spectrum of the curvaton field, $t_\sigma=d \ln {\cal P}_\sigma / d \ln k$.
The assumption that $\bar \sigma = 0$ will have real sense if the scale of interest,
$\ell_R=a_i/k_R$, will be larger than $\ell_c$ (both scales are calculated at
the end of inflation). The value of $\ell_R$ is given by the expression
\begin{equation}
\label{ellR}
\ell_R = \frac{a_i}{k_{end}}e^N.
\end{equation}
Here, $k_{end}$ is the scale leaving the horizon at the end of inflation, $a_i$ is
the scale factor at the end of inflation (and at the beginning of radiation era),
$N$ is a number of e-folds after the scale $k_R$ leaves the horizon. The condition
\begin{equation} \label{eq4}
\ell_c \ll \ell_R \ll \frac{a_i}{H_0}
\end{equation}
leads to the inequality $N\gg 1/t_\sigma$. It means that if
$t_\sigma$ is not small ($t_\sigma\sim 1$), and the coherent length $\ell_c$ is small,
one anticipates the blue curvature spectrum (the curvaton contribution)
and large non-Gaussianity {\it at small scales}.
In this case, the data at cosmological scales are described by the inflaton
fluctuations only.
In the opposite case, if $t_\sigma$ is very small,
the number of e-folds $N$, which is necessary for the fulfilment of the
condition  $\ell_R \gg \ell_c$ becomes large, $N\to N_{infl}\sim 60$.
In particular, if $t_\sigma \approx 1/60$, one has, instead of
the inequality (\ref{eq4}),
\begin{equation} \label{eq5}
\ell_c \sim \ell_R \sim \frac{a_i}{H_0}.
\end{equation}

Traditionally, predictions for the primordial curvature perturbation spectrum in a region
of small scales are constrained with a help of primordial black holes (PBHs).
PBHs are produced in the early Universe, e.g., in radiation era, due to
collapses of primordial density inhomogeneities
\cite{ZelNo, Hawking:1971ei, Hawking74, Carr:1974nx, Carr:1975qj, Polnarev:1986bi, Khlopov:2008qy}.
Experimental limits from PBH overproduction had been studied in many
articles, beginning from pioneering works \cite{Page:1976wx, Zeldovich1977};
for the latest reviews, see \cite{Josan:2009qn, Carr:2009jm}.

In the concrete case of the curvaton model, the idea of PBH constraining
at small scales was suggested in \cite{Lyth:2006gd} and was considered,
in more detail, in \cite{Kohri:2007qn}.

In the present work we consider the predictions of the mixed curvaton-inflaton scenario
just for the case which is most relevant for the PBH constraining:
we assume that {\it i)} the average value of the curvaton field in the Universe is zero,
and the Eq. (\ref{sigma2}) holds, and {\it ii)} the spectral tilt $t_\sigma$
is relatively large ($t_\sigma \sim 1$) and positive. In this case adiabatic
perturbations at small scales are produced mostly by the curvaton, resulting in a blue curvature spectrum.
Large non-Gaussianity follows in this scenario from the quadratic dependence of
the curvature on the curvaton field value.
In this case, the typical size of the ``curvaton domain'' \cite{Linde:2005yw}
is relatively small, it is smaller than the horizon size at the moment of the
formation of PBH with a given mass.

Recently, the PBH formation in a curvaton scenario was studied
in \cite{Kawasaki:2012wr, Kohri:2012yw}. In contrast with the present
work, authors of \cite{Kawasaki:2012wr, Kohri:2012yw} do not use the assumption
about a long period of inflation happened well before the observable Universe
left the horizon. They assume, instead, that the curvaton field is nearly
homogeneous in the whole Universe. The possibility of an essential PBH
production at small scales in such models depends on the concrete
inflationary scenario used. The authors of \cite{Kawasaki:2012wr} use for a
curvaton field a variant of the axion model
suggested in \cite{Kasuya:2009up} which predicts extremely
blue spectrum of curvature fluctuations, while the authors of \cite{Kohri:2012yw}
used the model with a convex potential [as the concrete realization of a
``hilltop curvaton'' scenario (see, e.g., \cite{Matsuda:2007av})], in which
strong scale dependence of the curvature power spectrum arises due to
tachyonic enhancement effects.

The plan of the paper is as follows. In the next Section we derive the
basic formula for the curvature perturbation spectrum used in the concrete
calculations. In Sec. \ref{sec-pbh-prod}, the process of PBH production
in our curvaton model is considered. The last Section contains the
results of the calculation and conclusions. The technical details
concerning the calculation of a probability density function (PDF) of the smoothed
curvature field are discussed in the Appendix.

\section{Curvature perturbation spectrum formula}
\label{zeta}

Calculations of primordial curvature power spectra in mixed curvaton-inflaton scenario
are carried out, in most cases, using the separated universe assumption and
$\delta N$-formalism
\cite{Starobinsky:1982ee, Starobinsky:1986fxa, Salopek:1990jq,
Sasaki:1995aw, Sasaki:1998ug, Wands:2000dp, Lyth:2005fi, Lyth:2004gb}.
It had been shown, in particular \cite{Lyth:2004gb}, that the
nonlinear curvature perturbation on an uniform energy density
hypersurface, given by the formula
\begin{equation}
\label{zt}
\zeta({\bf x}) = \psi(t, {\bf x}) + \frac{1}{3}
\int\limits_{\bar \rho(t)}^{\rho(t, {\bf x})} \frac{d\tilde\rho}{\tilde P + \tilde\rho},
\end{equation}
is conserved on superhorizon scales, for a fluid with an equation of state $P=P(\rho)$.
In Eq. (\ref{zt}), $\psi$ is the ``nonlinear curvature perturbation'' entering
the expression for the locally defined scale factor
\begin{equation}
a({\bf x}, t) = a(t) e^{\psi(t, {\bf x})}.
\end{equation}
In our case there are two (non-interacting) fluids, radiation from an inflaton
decay and an oscillating curvaton which we consider as pressureless matter
field. Assuming that the curvaton decays on an uniform total density
hypersurface, one has $\psi = \zeta$ on this surface, and, from Eq. (\ref{zt}),
one has
\begin{equation}
\zeta_r = \zeta+ \frac{1}{4} \ln \frac {\rho_r}{\bar \rho_r},
\end{equation}
\begin{equation}
\zeta_\sigma = \zeta + \frac{1}{3} \ln \frac {\rho_\sigma}{\bar \rho_\sigma}.
\end{equation}
From here, one has for the fluid densities
\begin{equation}
\label{rs000}
\rho_\sigma = \bar \rho_\sigma e^{3(\zeta_\sigma-\zeta)}, \qquad
\rho_r = \bar \rho_r e^{4(\zeta_r-\zeta)}.
\end{equation}
In the sudden decay approximation \cite{Lyth:2001nq, Malik:2002jb, Gupta:2003jc},
the sum of densities is, on the decay hypersurface, equal to $\bar \rho(t_{dec})$
(i.e., it is homogeneous quantity). It leads to the important relation \cite{Sasaki:2006kq}
\begin{equation} \label{eq11}
(1-\Omega_{\sigma, dec}) e^{4(\zeta_r-\zeta)} + \Omega_{\sigma, dec}e^{3(\zeta_\sigma-\zeta)} = 1,
\end{equation}
\begin{equation}
\Omega_{\sigma, dec} =\left. \frac{\bar \rho_\sigma}{\bar \rho_\sigma + \bar \rho_r}\right| _{dec}.
\end{equation}
The second relation which is necessary for the calculation of the curvature power spectrum
is the nonlinear generalization of the formula for the relative entropy perturbation.
In linear theory, one has
\begin{equation}
S_{\sigma r} = 3 (\zeta_\sigma - \zeta_r) = - 3 H
\left( \frac{ \delta\rho_\sigma }{\dot\rho_\sigma} -
  \frac{\delta\rho_r}{\dot \rho_r} \right).
\end{equation}
Neglecting the curvaton density compared with radiation density (at the beginning of the
radiation era), one has
\begin{equation} \label{Srapp}
S_{\sigma r} = 3 (\zeta_\sigma - \zeta_r) \approx - 3 H \frac{ \delta\rho_\sigma }{\dot\rho_\sigma}.
\end{equation}
The nonlinear extension of Eq. (\ref{Srapp}) is given by
\begin{equation} \label{SrappL}
S_{\sigma r}  \approx \ln \frac{\rho_\sigma}{\bar \rho_\sigma}, \quad
\rho_\sigma \approx \bar \rho_\sigma e^{S_{\sigma r}}.
\end{equation}

Using Eqs. (\ref{eq11}, \ref{SrappL}) one can connect the curvature perturbation $\zeta$ with
the curvaton field value on super-Hubble scales during inflation. At a beginning of
the curvaton oscillations, one has, in a case of the quadratic potential
\begin{equation}
\label{rsm2}
\bar \rho_\sigma e^{S_{\sigma r}} = \frac{1}{2}m_{osc}^2\sigma_{osc}^2.
\end{equation}
Here, $m_{osc}$ is the curvaton mass at the moment of the beginning
of oscillations. For simplicity, everywhere below we neglect the change of curvaton mass
after $t=t_{osc}$, and put $m_{osc}\approx m$.

It is convenient to study the evolution of the curvaton field (from the field value at
horizon exit during inflation, $\sigma_*$, to the field value at the beginning
of the oscillations, $\sigma_{osc}$) separately for the averaged value and perturbation,
\begin{equation} \label{eq17}
\sigma_* =  \bar \sigma_* + \delta\sigma_*\;, \qquad \sigma_{osc}=\bar\sigma_{osc} + \delta\sigma_{osc}.
\end{equation}
The equations determining the evolution are
\begin{equation} \label{ddotsigma}
\ddot{ \bar \sigma} + 3 H(t) \dot {\bar\sigma} + V' = 0,
\end{equation}
\begin{equation} \label{ddotdeltasigma}
\ddot {\delta \sigma} + 3 H(t) \dot{\delta\sigma }+ V''\delta\sigma = 0
\end{equation}
(the prime and the dot denote $\frac{d}{d\sigma}$ and $\frac{d}{dt}$, respectively).
Eq. (\ref{ddotdeltasigma}) is written for perturbations on superhorizon
scales, where the gradient term ($\sim k^2/a^2$) is negligible.
For a quadratic potential $V$, a fractional perturbation, $\delta\sigma/{\bar \sigma}$,
remains constant during the evolution.

As is pointed out in the Introduction, we assume that the early Universe follows
the scenario considered in \cite{BD, Starobinsky:1982ee} (``the Bunch-Davies case'').
In this scenario, $\bar \sigma$ is close to zero. As for the $\delta\sigma_*$,
one can neglect its evolution {\it during inflation}. When the curvaton field is close to
a minimum of the potential, then, due to a competition between the random walk
and a (slow) roll, the typical value of the field, as can be easily shown,
is $\sim \frac{H^2}{2\pi m}$, which is consistent with Eq. (\ref{sigma2}).

After an end of inflation, the evolution of the total curvaton field takes place
(of the average value as well as of the perturbation). Following
Ref. \cite{Lyth:2003dt}, we denote this evolution introducing the notation
\begin{equation}
\bar\sigma_{osc} = g(\bar \sigma_e), \qquad \delta\sigma_{osc} = g(\delta \sigma_*),
\end{equation}
where $\bar \sigma_e$ is the average value of the curvaton field at the end of inflation,
\begin{equation}
\bar \sigma_e = \bar \sigma_* e ^ {-\frac{1}{2}N t_\sigma}
\end{equation}
(which will be put equal to zero in final formulas). In a case of the quadratic potential
the evolution is linear, so
\begin{equation}
g(\delta\sigma_*) = g' \delta\sigma_* \;, \qquad g'=\frac{\delta\sigma_{osc}}{\delta\sigma_*},
\end{equation}
and one has, finally,
\begin{equation}
\sigma_{osc} = g(\bar\sigma_e) + g' \delta\sigma_*\;,
\end{equation}
\begin{equation}
g(\bar\sigma_e) = g' \bar\sigma_e \equiv \bar g = \bar\sigma_{osc}.
\end{equation}

The following steps are straightforward (see, e.g., \cite{Langlois:2008vk, Fonseca:2012cj}.
The entropy perturbation $S_{\sigma r}$ is obtained from Eq. (\ref{rsm2}), expanding
left and right sides of it up to second order,
\begin{equation}
S_{\sigma r} = 2 \frac{g'}{\bar g}\delta\sigma_* - \frac{g'^2}{{\bar g}^2}(\delta\sigma_*)^2.
\end{equation}
Further, expanding exponents in Eq. (\ref{eq11}) up to second order, one obtains, using
the connection of $S_{\sigma r}$ with $\zeta_\sigma$, $\zeta_r$:
\begin{eqnarray} \label{zeta2terms}
\zeta = \zeta_r + \frac{2}{3} \frac{g'}{\bar g} R_{\sigma,dec} \delta\sigma_* +
\qquad\qquad\qquad\qquad\qquad\qquad \\ \nonumber
+\; \frac{2}{9}\left[ \frac{3}{2} R_{\sigma,dec} - 2R_{\sigma,dec}^2 - R_{\sigma,dec}^3\right]
\left( \frac{g'}{\bar g} \right)^2 \delta\sigma_*^2.
\end{eqnarray}
Here, $R_{\sigma,dec}$ is given by the formula
\begin{equation}
\label{Rsigmadecay}
R_{\sigma,dec} = \frac{3 \Omega_{\sigma, dec}}{4 - \Omega_{\sigma, dec}}.
\end{equation}
Since, according to the definition of the $R_{\sigma,dec}$,
there is the proportionality $R_{\sigma,dec} \sim \bar{\rho_\sigma} = \frac{1}{2}m\bar \sigma_{osc}^2$
(the proportionality coefficient is derived below, in Sec. \ref{sec-pbh-prod}-A), and
since $\bar g = \bar \sigma_{osc}$, it follows from Eq. (\ref{zeta2terms}) that
only the term proportional to $R_{\sigma,dec}/{\bar g}^2$ survives in this Equation in the limit
$\bar \sigma_{osc} \to 0$. It leads to the simple formula for the curvaton-generated part of the total
curvature perturbation:
\begin{equation}
\label{eq28zeta}
\zeta - \zeta_r \equiv \zeta_{(\sigma)} =
\frac{1}{3} R_{\sigma,dec} \left( \frac{g'}{\bar g} \right)^2 (\delta\sigma_*)^2.
\end{equation}
Everywhere below we will use for $\zeta_{(\sigma)}$ the notation $\zeta_{\sigma}$,
dropping the brackets in the index.

The power spectrum of $(\delta\sigma_*)^2$ is expressed through the power spectrum
of the curvaton field perturbation \cite{Lyth:2006gd},
\begin{equation}
{\cal P}_{\delta\sigma_*^2}^{1/2} = \left( \frac{4}{t_\sigma} {\cal P}_{\sigma_*}^2\right)^{1/2},
\end{equation}
and the power spectrum of the curvaton field is
\begin{equation}
{\cal P}_{\sigma_*}= \left( \frac{H_i}{2\pi}\right)^2
\left( \frac{k}{k_R}\right)^{t_\sigma} =
\left( \frac{H_i}{2\pi}\right)^2 e^{-(N_{infl} - N)t_\sigma} \left( \frac{k}{H_0}\right)^{t_\sigma}.
\end{equation}
The spectral tilt $t_\sigma$ is simply connected with a value of the effective mass
of the curvaton field, $m_*$:
\begin{equation}
t_\sigma = \frac{2 m_{\sigma*}^2}{3 H_i^2}.
\end{equation}
The difference $N_{infl} - N$ is the number of e-folds of ``relevant inflation''  \cite{Lyth:2006gd},
i.e., the number of e-folds passed from the moment when the observable Universe leaves horizon
up to the moment when the scale $k_R^{-1}$ leaves horizon. The scale $k_R^{-1}$ enters
horizon at the radiation era, just when the curvature perturbation $\zeta_{\sigma}$ is
created. The value of $k_R$ determines the value of horizon mass $M_h$ and, correspondingly,
the order of magnitude value of PBH mass that can be produced at this moment.

Finally, we obtain for the curvature spectrum the expression
\begin{eqnarray}
\label{Pzetafinal}
{\cal P}_{\zeta_\sigma}^{1/2} = \frac{2}{3}R_{\sigma,dec} \frac{g'^2}{\bar g^2} \frac{1}{\sqrt{t_\sigma}}
\frac{H_i^2}{(2\pi)^2}
\left( \frac{k}{k_R}\right)^{t_\sigma}.
\end{eqnarray}

For calculations using this formula, one needs the relation $R_{\sigma,dec}/ \bar g^2$.
It is derived in the next Section, for the concrete choice of the potential
[see Eq. (\ref{RsigmaOVERg2})].

\section{PBH production in the curvaton model}
\label{sec-pbh-prod}

\subsection{Curvaton potential}

Recently, a variety of models of chaotic inflation in supergravity, in connection with the
curvaton scenario and curvaton web problem, had been introduced and studied
\cite{Demozzi:2010aj}. Their models and conclusions,
however, can not be used in our work straightforwardly because in our
curvaton scenario {\it i)} there is no degeneracy of masses of the inflaton
and curvaton fields, and {\it ii)} our curvaton field is a real, single component
field, rather than the radial component of a complex field, as in \cite{Demozzi:2010aj}.
Both these features are not inconsistent with the general theory of chaotic inflation
in supergravity \cite{Kallosh:2010ug, Kallosh:2010xz}: for example, the
curvaton field can be imaginary part of the complex scalar field \cite{Kallosh:2010xz}.

We consider the model with the simple phenomenological potential of the form
\begin{equation}
\label{Vsigma}
V(\sigma) = \frac{\sigma^2}{2}\left(m^2 + \alpha H^2(t)\right).
\end{equation}
The corresponding effective mass of the curvaton field is $m_*^2 = m^2 + \alpha H^2$
and the spectral tilt is given by
\begin{equation}
t_\sigma = \frac{2}{3}\left( \alpha + \frac{m^2}{H_i^2}\right)
 \approx \frac{2}{3}\alpha.
\end{equation}

The evolution equation for the curvaton field $\delta\sigma$
is given above [see Eq. (\ref{ddotdeltasigma})].
The calculation of $\delta\sigma(t)$ starts at moment $t=0$
corresponding to an end of inflation and the beginning
of the radiation-dominated era (the reheating is assumed to be instant).

The derivative $g'$ is calculated numerically, and the initial conditions are:
\begin{equation}
\delta\sigma(t=0) = \delta\sigma_*, \qquad \dot {\delta\sigma}(t=0) = 0.
\end{equation}
In our case, because the potential (\ref{Vsigma}) is quadratic,
$g'=\delta\sigma_{osc}/\delta\sigma_{*}$. For the value of $\delta \sigma_{osc}$, we take
$\delta \sigma_{osc} \equiv \delta \sigma(t_{osc})$, and the moment of time
when oscillations start, $t_{osc}$, is determined by the condition \cite{Kawasaki:2011pd}
\begin{equation}
\label{tosc-condition}
\left| \frac{ \delta\sigma}{ \dot {\delta \sigma} } \right|_{t_{osc}} = H(t_{osc})^{-1}.
\end{equation}
According to this condition, after an onset of the oscillation the
time scale of a change of the curvaton field is
smaller that the expansion time $H^{-1}$.

\begin{figure}
\center %
\includegraphics[width=8.5 cm, trim = 0 6 0 0 ]{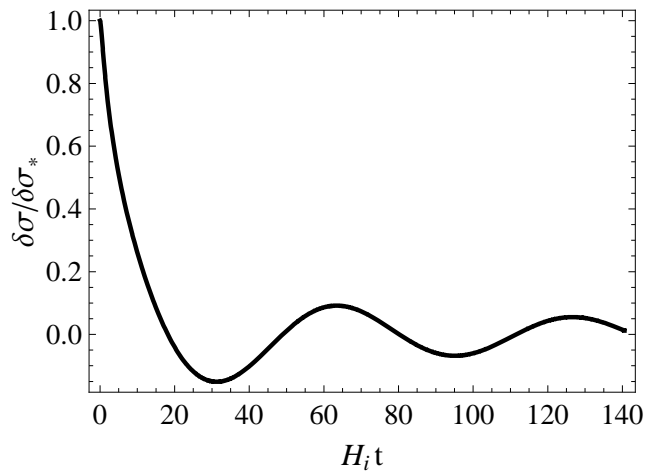}
\caption{ \label{fig-sse}
The solution of Eq. (\ref{ddotdeltasigma}) for $\delta\sigma(t)$, for
$m = 0.1 H_i$, $\alpha=1$.
}
\end{figure}

The example of the solution of Eq. (\ref{ddotdeltasigma}) for the particular set
of parameters, $m/H_i = 0.1$, $\alpha=1$, is shown in Fig. \ref{fig-sse}.
The corresponding value of the derivative $g'$ is equal to $0.62$.

According to (\ref{Vsigma}), the energy density of the (average) curvaton field at the moment $t_{osc}$
is [$H_{osc}\equiv H(t_{osc})$]
\begin{eqnarray}
\label{eqR1}
\bar\rho_{\sigma,osc} = \frac{\bar \sigma_{osc}^2}{2} (m^2 + \alpha H_{osc}^2).
\end{eqnarray}
After the moment $t=t_{osc}$, and until the curvaton's decay at $t=t_{dec}$, the curvaton
is assumed to behave like a pressureless matter, so a value of the curvaton density at decay time is
[$a(t_{osc})\equiv a_{osc}$, $a(t_{dec})\equiv a_{dec}$]:
\begin{eqnarray}
\label{eqR2}
\bar\rho_{\sigma,dec}=\bar\rho_{\sigma,osc} \left( \frac{a_{osc}}{a_{dec}} \right)^3.
\end{eqnarray}
The radiation density at the moment $t_{dec}$ can be related to $H(t_{dec})\equiv H_{dec}$
by using the Friedmann equation,
\begin{eqnarray}
\label{eqR3}
H_{dec}^2 = \frac{8 \pi}{3 m_{Pl}^2} \bar \rho_{r,dec}
\end{eqnarray}
(here and below we neglect $\bar\rho_{\sigma,dec}$ compared to $\bar\rho_{r,dec}$).
From Eqs. (\ref{eqR1}, \ref{eqR2}, \ref{eqR3}) one obtains
\begin{eqnarray}
\label{Omegasigmadec}
\Omega_{\sigma,dec} = \frac{\bar\rho_{\sigma,dec}}{\bar\rho_{r,dec}} =
\frac{4\pi}{3 m_{Pl}^2} \bar \sigma_{osc}^2
\left( \alpha + \frac{m^2}{H_{osc}^2} \right)   \frac{a_{dec}}{a_{osc}}.
\end{eqnarray}
Now, from Eqs. (\ref{Rsigmadecay}, \ref{Omegasigmadec}), taking into account that
$\Omega_{\sigma,dec} \ll 1$ and using relations $a \sim t^{1/2} \sim H^{-1/2}$, we obtain the final formula used in our calculations,
\begin{eqnarray}
\label{RsigmaOVERg2}
\frac{R_{\sigma,dec}}{\bar g^2} = \frac{\pi}{m_{Pl}^2 \sqrt { 2 \Gamma_\sigma t_{osc}} }
\left( \alpha + \frac{m^2}{H(t_{osc})^2} \right).
\end{eqnarray}
In this Equation, we used the equality $\frac{1}{2 t_{dec}} = H_{dec} = \Gamma_\sigma$,
to obtain $t_{dec}$, while $t_{osc}$ is calculated numerically from the condition
given by Eq. (\ref{tosc-condition}).

Note also that in a case when $\alpha=0$ and $H_{osc}=m$, one obtains from Eq. (\ref{Omegasigmadec})
\begin{eqnarray}
\Omega_{\sigma,dec} = \frac{1}{6} \left( \frac{\bar \sigma_{osc}}{M_P} \right)^2 \sqrt{\frac{m}{\Gamma_\sigma}}
\end{eqnarray}
($M_P=m_{Pl}/\sqrt{8\pi}$), which corresponds to a well-known result (see, e.g., \cite{Lyth:2001nq}).

\subsection{PDF for the curvature perturbation $\zeta$}

It is generally assumed that the perturbations of the curvaton field at Hubble exit
during inflation can be well described by a Gaussian random field (correspondingly,
the equation (\ref{eq17}) for $\sigma_*$ contains, in its right-hand side,
no higher-order terms). In our curvaton model, the curvature perturbation $\zeta_\sigma$
depends on the curvaton field quadratically. In this case, the field $\zeta_\sigma$ is
chi-squared distributed, so the probability density function for $\zeta_\sigma$
perturbations is strongly non-Gaussian.

A formula for the PDF in the case of chi-square distribution of $\zeta$-filed perturbations,
i.e., in the case when
\begin{equation} \label{zsPDF}
\zeta_\sigma({\bf x}) = A\left[ \chi({\bf x})^2 - \langle \chi^2\rangle \right]
\end{equation}
is well known \cite{Matarrese:2000iz} (in our notations, $\chi\equiv\delta\sigma_*$;
in contrast with the analogous formula (\ref{eq28zeta}) in Sec. \ref{zeta}, in Eq. (\ref{zsPDF})
the subtraction of $\langle \chi^2\rangle$ is performed, to provide the condition
$\langle \zeta_\sigma \rangle=0$).

For applications in PBH production calculations (with using the Press-Schechter
formalism \cite{PS}) one must derive the PDF for the {\it smoothed} field $\zeta_\sigma$.
This problem is thoroughly discussed in the Appendix. It is argued there
that the PDF for the smoothed $\zeta$ field can be approximately written
in the form
\begin{equation} \label{pzRnu}
p_{\zeta, R}(\zeta_R) \approx \frac{1}{\sigma_\zeta(R)} p(\tilde \nu), \quad
\tilde\nu \equiv \frac{\zeta_R}{\sigma_\zeta(R)}.
\end{equation}
Here, $\sigma_\zeta(R)$ is the variance of the smoothed $\zeta$ field [it is
given by Eq. (\ref{A34})] and the function $p(\tilde \nu)$ is given by Eq. (\ref{A10}).
Effects of the smoothing operation enter, in Eq. (\ref{pzRnu}), only through
the variance, while the function $p(\tilde \nu)$ is the same in smoothing and
non-smoothing cases.

\begin{figure}
\center %
\includegraphics[width= 8.5 cm, trim = 0 0 0 0 ]{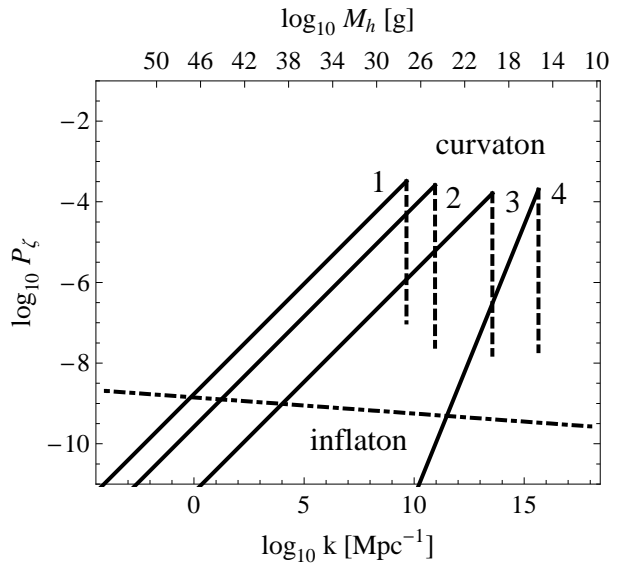}
\caption{ \label{fig-power-sp}
The examples of curvature perturbation power spectrum ${\cal P}_\zeta(k)$ calculation for the curvaton
model considered.
Curve 1 - $H_i=10^{12.5}\;$GeV, $\Gamma_\sigma/m=10^{-24.8}$, $\alpha=0.4$;
curve 2 - $H_i=10^{13}\;$GeV, $\Gamma_\sigma/m=10^{-22.7}$, $\alpha=0.4$;
curve 3 - $H_i=10^{14}\;$GeV, $\Gamma_\sigma/m=10^{-18.5}$, $\alpha=0.4$;
curve 4 - $H_i=10^{15}\;$GeV, $\Gamma_\sigma/m=10^{-15.3}$, $\alpha=1$.
For all cases, $m=0.1 H_i$.
For reference, the curvature perturbation
power spectrum generated by the inflaton is also shown, assuming the spectral
index $n_i=0.96$ has zero running on cosmological as well as smaller scales.
}
\end{figure}

\subsection{PBH mass spectrum and constraints}

The PBH constraints are obtained using the Press and Schechter formalism
generalized for a case of non-Gaussian PDFs. We will follow the Refs.
\cite{Bugaev:2011wy, Lyth:2012yp, Byrnes:2012yx, Linde:2012bt} working
with the curvature perturbation $\zeta_R$ rather than with the density
contrast. The basic formula in the Press and Schechter approach is
\begin{eqnarray}
\label{PSformalism}
\frac{1}{\rho_i} \int\limits_M^{\infty} \tilde M n(\tilde M) d \tilde M = \nonumber
\qquad \qquad \qquad \qquad \qquad \qquad \\ = \int\limits_{\zeta_{c}}^{\infty}
p_{\zeta,R}(\zeta_R) d\zeta_R = P(\zeta_R>\zeta_{c}; R(M), t_i).
\end{eqnarray}
In this Equation, $P$ is the probability that in a region of comoving size $R$
one has $\zeta_R > \zeta_c$, where $\zeta_c$ is the threshold value for the
PBH formation in the radiation era, $n(M)$ is the mass spectrum of the
collapsed objects, $\rho_i$ is the initial energy density. We will use the
value of $\zeta_c=0.75$ corresponding to the PBH formation criterion for
the density contrast, $\delta_c=1/3$.

The PBH mass $M_{BH}$ is connected with the mass of the fluctuation $M$ by
the relation \cite{Bugaev:2000bz, Bugaev:2008gw}
\begin{eqnarray} \label{MBHMh23}
M_{BH} \cong f_h M_h = f_h M_i^{1/3} M^{2/3},
\end{eqnarray}
where $M_h$ is the horizon mass corresponding to the time when the fluctuation
of mass $M$ crosses horizon in radiation era, $M_i$ is the horizon mass
at the start of the radiation era, $t=t_i$.
For the constant $f_h$ we will use the value $f_h=(1/3)^{1/2}$ \cite{Bugaev:2000bz, Bugaev:2008gw}.
In the approximation of the fast
reheating, $t_i$ coincides with the time of the end of inflation.

\begin{figure}[!t]
\center %
\includegraphics[width= 8.2 cm, trim = 0 0 0 0 ]{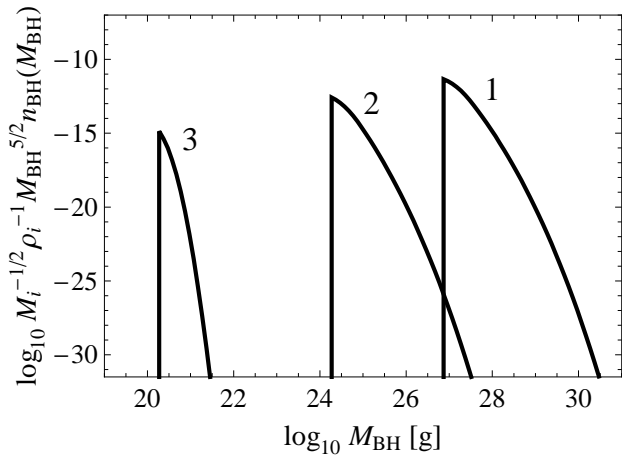}
\caption{ \label{fig-nbh}
Examples of the PBH mass spectra calculations.
Curve 1 - $H_i=10^{12.5}\;$GeV, $\Gamma_\sigma/m=10^{-24.8}$, $\alpha=0.4$;
curve 2 - $H_i=10^{13}\;$GeV, $\Gamma_\sigma/m=10^{-22.7}$, $\alpha=0.4$;
curve 3 - $H_i=10^{14}\;$GeV, $\Gamma_\sigma/m=10^{-19.7}$, $\alpha=1$;
For all cases, $m=0.1 H_i$, $\zeta_c=0.75$.
}
\end{figure}

Using Eqs. (\ref{PSformalism}) and (\ref{MBHMh23}) one obtains
the formula for the PBH number density (mass spectrum) \cite{Bugaev:2011wy}:
\begin{equation}
\label{nBH}
n_{BH}(M_{BH}) = \left( \frac{4 \pi}{3} \right)^{-1/3}
\left| \frac{\partial P}{\partial R }\right|
 \frac{f_h \rho_i^{2/3} M_i^{1/3} } {a_i M_{BH}^2},
\end{equation}
where $a_i$ is the scale factor at the end of inflation,
\begin{equation}
a_i = \frac{a_{eq}}{\sqrt{2} H_i^{1/2} t_{eq}^{1/2}},
\end{equation}
and $a_{eq}, t_{eq}$ are scale factor and time at matter-radiation equality, respectively.
The derivative $\partial P/ \partial R$ is given by the expression
\begin{equation}
\frac{\partial P}{\partial R } = \frac{\zeta_c}{\sigma_\zeta(R)}
 \frac{d \sigma_\zeta(R)}{dR} p_{\zeta,R}(\zeta_c).
\end{equation}
This expression is obtained with using the formula (\ref{pzRnu}) for the non-Gaussian PDF.
The dependence of the PBH number density on the curvature perturbation power spectrum
${\cal P}_\zeta$ arises just through the derivative $\partial P/ \partial R$.

If PBHs form at $t=t_e$, one can calculate the energy density fraction of the Universe
contained in PBHs at the time of formation (at this time, the horizon mass is equal to
$M_h(t_e) \equiv M_h^f$ \cite{Bugaev:2011wy}):
\begin{eqnarray} \label{qq1}
\label{omPBH-beta}
\Omega_{PBH}(M_h^f) \approx \nonumber \qquad\qquad\qquad\qquad\qquad\qquad\qquad\qquad\qquad \\
\approx \frac{1}{\rho_i} \left( \frac{M_h^f}{M_i} \right)^{1/2} \int n_{BH}(M_{BH}) M_{BH}^2 d \ln M_{BH}
\approx \nonumber  \\ \approx
\frac{(M_h^f)^{5/2}}{\rho_i M_i^{1/2}} n_{BH}( M_{BH}) \left. \right|_{ M_{BH} = M_{BH}^{min}} . \qquad
\end{eqnarray}
In this formula, $M_{BH}^{min}$ is the minimum mass of the PBH mass spectrum,
$M_{BH}^{min} \approx f_h M_h^f$. The PBH mass spectrum is very steep, so, with high
accuracy one has
\begin{eqnarray} \label{qq2}
\Omega_{PBH}(M_h^f) \approx \beta_{PBH}(M_h^f),
\end{eqnarray}
where $\beta_{PBH}$ is, by definition (see, e.g., \cite{Carr:2009jm}), the fraction
of the Universe's mass in PBHs at their formation time,
\begin{eqnarray}
\beta_{PBH}(M_h^f) \equiv \frac{\rho_{PBH}(t_e)}{\rho(t_e)}.
\end{eqnarray}
Now, having Eqs. (\ref{qq1}, \ref{qq2}), one can use the experimental limits on the value of
$\beta_{PBH}$ \cite{Carr:2009jm} to constrain parameters of models used for
PBH production predictions.

\section{Results and discussion}
\label{sec-results}

The examples of curvaton-generated curvature perturbation power spectra are
shown in Fig. \ref{fig-power-sp}, and some examples of the PBH mass spectra calculations
are given in Fig. \ref{fig-nbh}. For each curve shown in Figs. \ref{fig-power-sp}, \ref{fig-nbh},
the model parameter $\Gamma_\sigma$ is chosen so that the predicted PBH abundance
is of the same order of magnitude as the currently available limits \cite{Carr:2009jm} on the
parameter $\beta_{PBH}$ in the corresponding PBH mass range.
On the vertical axis of Fig. \ref{fig-nbh} the combination
$M_i^{-1/2} \rho_i^{-1} M_{BH}^{5/2} n_{BH}(M_{BH})$ is shown; just this combination
is approximately equal to $\beta_{PBH}$, as it follows from Eq. (\ref{omPBH-beta}).

\begin{figure}[!t]
\center %
\includegraphics[width=7.6 cm, trim = 0 0 0 0 ]{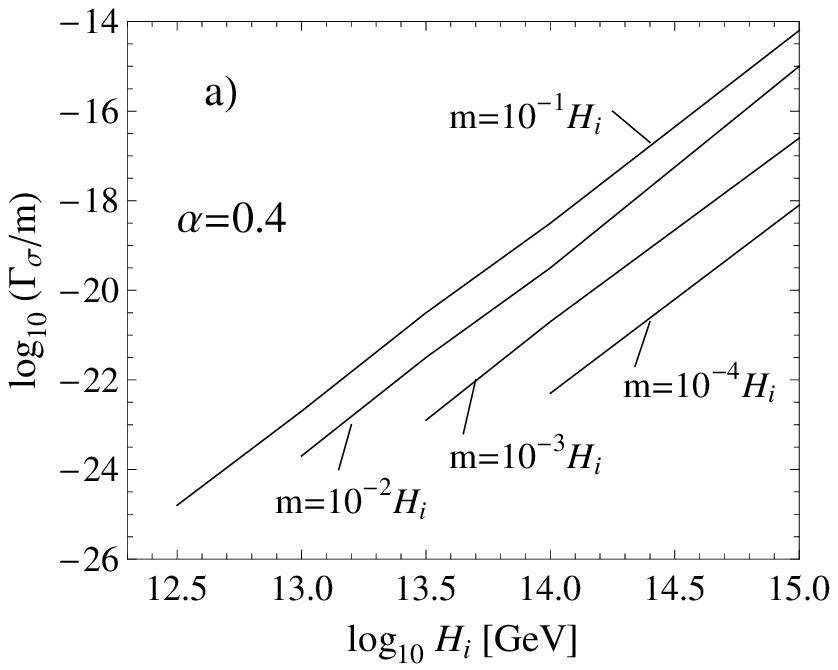}
\includegraphics[width=8.6 cm, trim = 0 0 0 0 ]{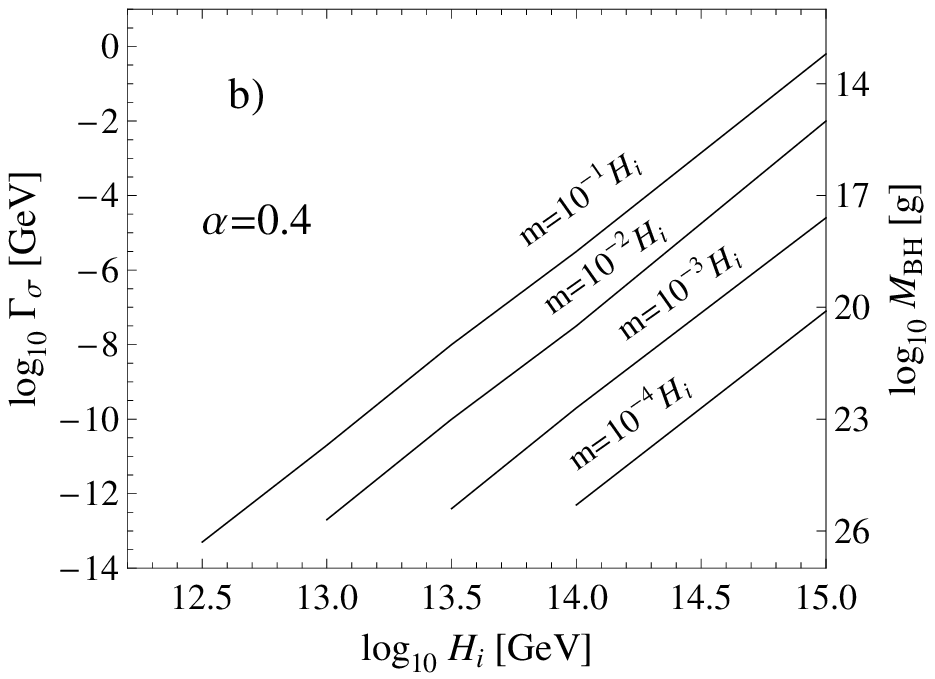}
\includegraphics[width=7.6 cm, trim = 0 0 0 -20 ]{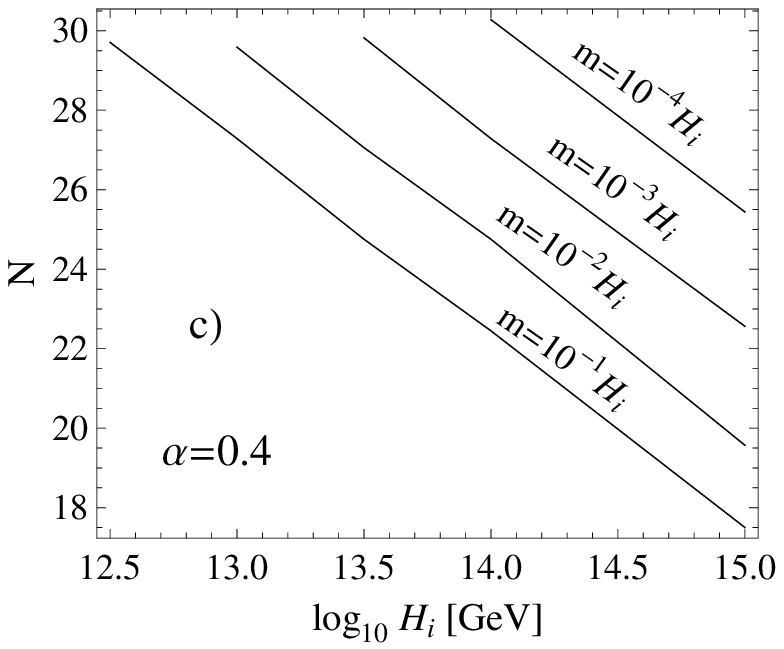}
\caption{ \label{fig-Gsm-04}
a), b) The resulting constraints on the values of model parameters obtained for the
curvaton model considered in this paper (for $\alpha=0.4$).
Regions below the lines correspond to the sets of parameters that are prohibited
by PBH overproduction.
c) The values of $N$ corresponding to the constraints,
as functions of Hubble parameter during inflation.
}
\end{figure}

The following connection between the comoving scale $k_R$ and horizon mass
$M_h$ (which is approximately equal to PBH mass) is used
in Fig. \ref{fig-power-sp} \cite{Bugaev:2010bb}:
\begin{equation}
k_R \approx \frac{2 \times 10^{23}}{\sqrt{M_{h} / 1 {\rm g}}} {\rm Mpc}^{-1}.
\end{equation}

It is seen from Fig. \ref{fig-nbh} that for smaller values of $\alpha$,
the PBH mass spectra become more wide. The low mass cut-off of the curves shown is
determined by the fact that no PBHs are formed before the curvaton decays at $t=t_{dec}$,
so the minimal PBH mass is $M_{BH}^{min}=f_h M_h(t_{dec})$.

\begin{figure}[!t]
\center %
\includegraphics[width=7.6 cm, trim = 0 0 0 0 ]{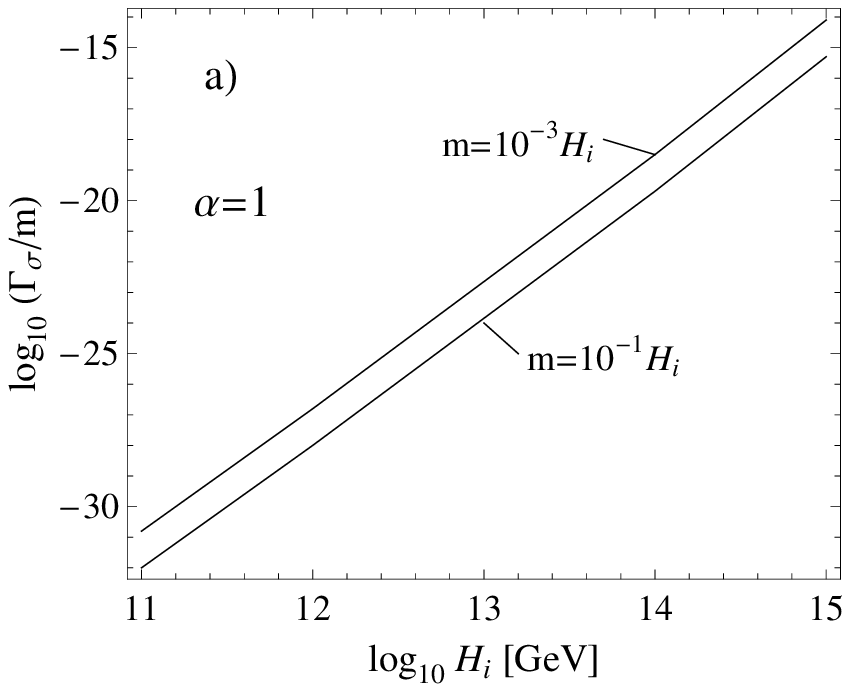}
\includegraphics[width=8.6 cm, trim = 0 0 0 0 ]{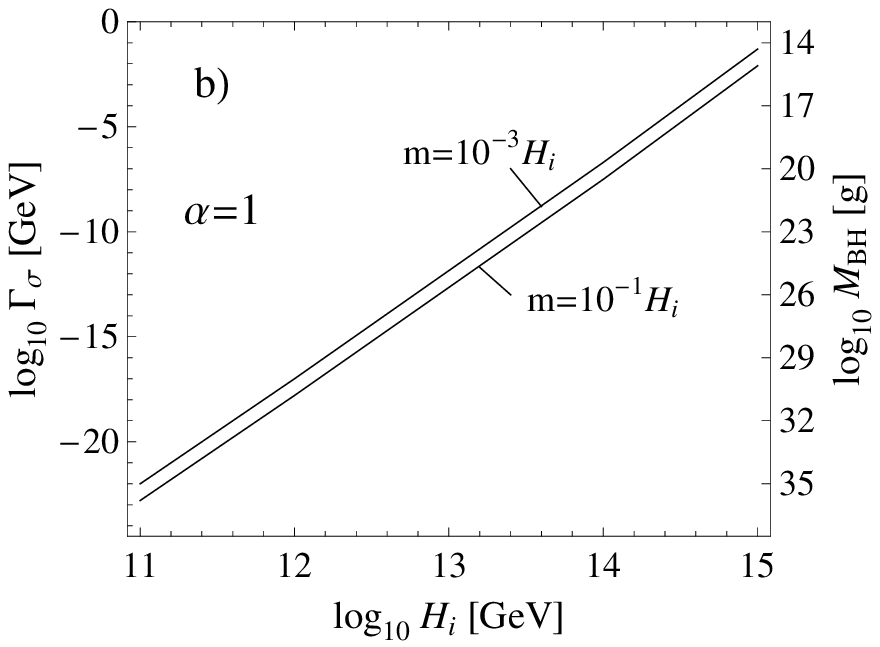}
\includegraphics[width=7.6 cm, trim = 0 0 0 -20 ]{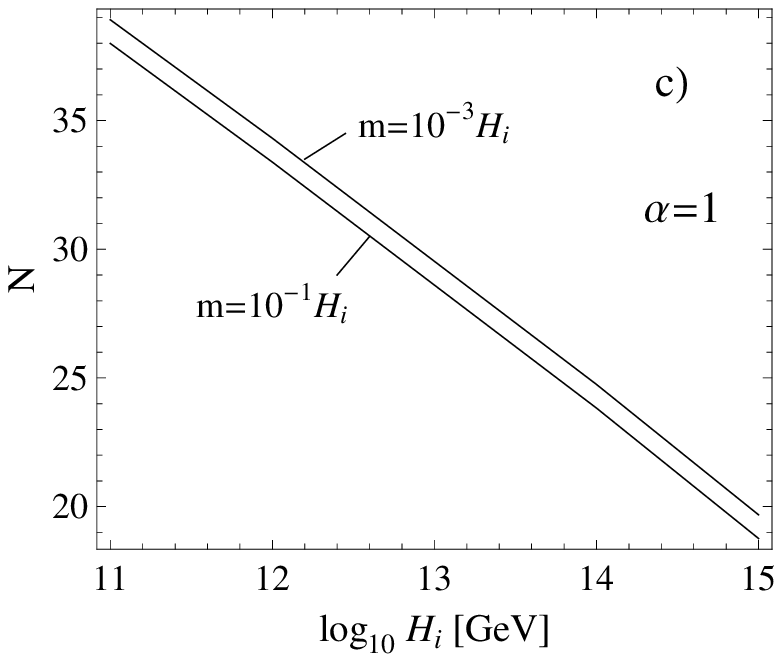}
\caption{ \label{fig-Gsm-1}
a), b) The resulting constraints on the values of model parameters obtained for the
curvaton model considered in this paper (for $\alpha=1$).
Regions below the lines correspond to the sets of parameters that are prohibited
by PBH overproduction.
c) The values of $N$ corresponding to the constraints,
as functions of Hubble parameter during inflation.
}
\end{figure}

For the constraining of the curvaton model parameters, we used the limits for
$\beta_{PBH}(M_{BH})$ from the review work \cite{Carr:2009jm}.
Demanding that PBHs are not overproduced, i.e., the value of
$\beta_{PBH}(M_{BH})$ does not exceed the available limits \cite{Carr:2009jm},
one may obtain the corresponding constraints on the parameters of the considered
cosmological model. Such constraints are shown in Fig. \ref{fig-Gsm-04} for
the case of $\alpha=0.4$ and in Fig. \ref{fig-Gsm-1} for $\alpha=1$.

In particular, in Figs. \ref{fig-Gsm-04}a and \ref{fig-Gsm-1}a we show the limits
on the combination of parameters $\Gamma_\sigma/m$ while in Figs. \ref{fig-Gsm-04}b
and \ref{fig-Gsm-1}b - on the value of $\Gamma_\sigma$ itself. The prohibited
(by PBH overproduction) parameter ranges lie below the corresponding lines.

In the sudden decay approximation, there is a very simple
approximate connection between $\Gamma_\sigma$ and the PBH mass produced.
It follows from the relations
\begin{equation}
\frac{M_h(t_{dec})}{M_i} = \frac{t_{dec}}{t_i} = \frac{H_i}{H_{dec}} = \frac{H_i}{\Gamma_\sigma},
\end{equation}
\begin{equation}
\label{MBH-GamSig}
M_{BH} \approx f_h M_h(t_{dec}) = \frac{f_h m_{Pl}^2}{16 \Gamma_\sigma} .
\end{equation}
Thus, constraints on $\Gamma_\sigma$ (see Figs. \ref{fig-Gsm-04}b,
\ref{fig-Gsm-1}b) are at the same time constraints
on the mass of PBHs that can be produced in this model [this is reflected on
the vertical axis of the Figures \ref{fig-Gsm-04}b, \ref{fig-Gsm-1}b; the relation between $M_{BH}$ and
$\Gamma_\sigma$ is given by Eq. (\ref{MBH-GamSig})].


Deriving the constraints, we use the condition
\begin{equation}
{\cal P}_{\zeta,\sigma} < 2.4 \times 10^{-9} \quad {\rm for } \quad k<k_c \approx 1\; {\rm Mpc}^{-1}
\end{equation}
in order not to contradict with the data on the cosmological scales.

One must note that the characteristic values of ${\cal P}_\zeta$ which determine the
constraints on the model parameters shown in Figs. \ref{fig-Gsm-04}, \ref{fig-Gsm-1} are
of order of $\sim 10^{-3.5}$. This is consistent with the PBH constraints on
${\cal P}_\zeta$ (for non-Gaussian $\zeta_\sigma$-perturbations) obtained
in our previous work \cite{Bugaev:2013vba} (see also \cite{Lyth:2012yp}).

In Figs. \ref{fig-Gsm-04}c, \ref{fig-Gsm-1}c we show also the number of e-folds after the scale $k_R$
leaves horizon,
\begin{equation}
N=\log \frac{k_{end}}{k_{R}} = \log \frac{a_i H_i}{a_{dec}H_{dec}} = \frac{1}{2}\log \frac{H_i}{\Gamma_\sigma},
\end{equation}
as a function of the constrained model parameters.
It is seen that for all cases corresponding to the obtained PBH constraints, $N \gg 1$. As
pointed out in the Introduction, this is needed for the validity of the considered model.

One can see from the resulting Figs. \ref{fig-Gsm-04}, \ref{fig-Gsm-1} that, generally, PBH constraints are
very weak. The forbidden region contains too small values of $\Gamma_\sigma/m$
(although the nucleosynthesis limit, $\Gamma_\sigma\gtrsim (1 {\rm MeV})^2/M_P$,
allows such values). The PBH constraint works only for very high values of Hubble constant
during inflation, $H_i\gtrsim 10^{11\div 12.5}\;$GeV, and for very large values of curvaton masses,
$m \gtrsim (10^{-4}\div 10^{-1}) H_i$. For other values of parameters, the spectrum amplitude,
${\cal P}_{\zeta_\sigma}$, is too small and cannot be constrained.
For illustrative purposes we show in Figs. \ref{fig-Gsm-04}, \ref{fig-Gsm-1} constraints
for a large interval of $H_i$ values, up to $10^{15}\;$GeV, although there is
a well-known upper bound on the Hubble parameter during inflation (according
to the recent results of {\it Planck} collaboration, $H_i/M_P < 3.7\times 10^{-5}$ \cite{Ade:2013uln}).
One must note also that in the forbidden region the reheating temperatures are rather high
($T_{RH}\sim \sqrt{H_i M_P}$) and, in standard supersymmetric models, gravitinos
are overproduced.

\section*{Acknowledgments}

The study was supported by The Ministry of education and science
of Russian Federation, project 8525.

\appendix \section{Moments of PDF of $\zeta$-field}

It follows from Eq. (\ref{eq28zeta}) that in our model the curvature perturbation depends on the
Gaussian curvaton field $\delta \sigma_*$ quadratically,
\begin{equation} \label{A1}
\zeta = A (\delta \sigma_* ^2 - \langle \delta \sigma_* ^2 \rangle),
\end{equation}
\begin{equation}
A \equiv \frac{1}{3} R_{\sigma,dec} \frac{g'^2}{{\bar g}^2}.
\end{equation}
In Eq. (\ref{A1}) the constant term $A \langle \delta \sigma_* ^2 \rangle$ is
subtracted such that now $\langle \zeta \rangle = 0$, and $\zeta$ is the ``overcurvature''.
Introducing the notation $\delta\sigma_* \equiv \chi$, one has
\begin{equation} \label{A3}
\zeta = A (\chi ^2 - \langle \chi ^2 \rangle),
\end{equation}
and the PDF of the $\chi$ field is
\begin{equation} \label{A4}
p_\chi(\chi) = \frac{1}{\sigma_\chi \sqrt{2\pi}} e^{-\frac{\chi^2}{2 \sigma_\chi^2}}, \qquad
\sigma_\chi^2 \equiv \langle \chi^2 \rangle.
\end{equation}
PDF of the $\zeta$ field is obtained from the PDF of the $\chi$ field using the
Chapman-Kolmogorov equation,
\begin{eqnarray} \label{A5}
p_\zeta(\zeta) = \int d\chi p_\chi(\chi) \delta_D \left[ \zeta - A (\chi ^2 - \langle \chi ^2 \rangle) \right]
= \nonumber \\ =
\int d\chi p_\chi(\chi) \sum\limits_i
\left[ \delta_D(\chi-\chi_i) \frac{1}{|\frac{d\zeta}{d\chi}(\chi_i)|}\right].
\end{eqnarray}
Here, $\chi_i$ are roots of the equation
\begin{equation}
A\chi^2 - A\langle \chi^2 \rangle - \zeta = 0.
\end{equation}
The final expression for the PDF of the $\zeta$ field is
\begin{equation}
\label{A7}
p_\zeta(\zeta) = \frac{1}{A \sqrt{\frac{\zeta}{A} + \langle \chi^2 \rangle} } \;
p_\chi\left(\sqrt{\frac{\zeta}{A} + \langle \chi^2 \rangle}\right).
\end{equation}
The variance of the $p_\zeta(\zeta)$ is
\begin{equation}
\langle \zeta^2 \rangle = \int\limits_{\zeta_{min}}^{\infty} \zeta^2 p_\zeta(\zeta) d\zeta =
2 \zeta_{min}^2= 2 A^2 \langle \chi^2 \rangle^2.
\end{equation}
Using this equation, the distribution (\ref{A7}) can be written in the form:
\begin{equation}  \label{A9}
p_\zeta(\zeta) = \frac{1}{\langle\zeta^2 \rangle^{1/2}} p(\nu),
\end{equation}
\begin{equation} \label{A10}
p(\nu) = \frac{1}{\sqrt{1 + \sqrt{2}\nu } } e^{-\frac{1}{2} (1 + \sqrt{2}\nu )}.
\end{equation}
In this equation, the notation $\nu = \zeta/\langle \zeta^2 \rangle^{1/2}$ is introduced.
Note, that the product $p_\zeta(\zeta)d\zeta$ doesn't depend on $\zeta$ and
$\langle \zeta^2 \rangle^{1/2}$ separately, i.e.,
\begin{equation}
p_\zeta(\zeta)d\zeta = p(\nu) d\nu.
\end{equation}

The first (central) moments of the $p_\zeta$ are
\begin{equation}
\langle \zeta^3\rangle = 8 A^3 \langle\chi^2\rangle^3, \quad
\langle \zeta^4\rangle = 60 A^4 \langle\chi^2\rangle^4,
\end{equation}
and the first {\it cumulants}, $\langle \zeta^n \rangle_c$, are
given by the relations (see, e.g., \cite{Smith95})
\begin{eqnarray} \label{A13}
\langle \zeta^2\rangle_c = \langle \zeta^2 \rangle, \quad
\langle \zeta^3\rangle_c = \langle \zeta^3 \rangle, \; \qquad \qquad \qquad  \\ \nonumber
\langle \zeta^4\rangle_c = \langle \zeta^4 \rangle - 3 \langle \zeta^2 \rangle^2, \quad
\langle \zeta^5\rangle_c = \langle \zeta^5 \rangle - 10 \langle \zeta^2 \rangle \langle \zeta^3 \rangle.
\end{eqnarray}
The reduced cumulants are defined by the relation (see, e.g., \cite{Scoccimarro:2000qg})
\begin{equation}
D_n \equiv \frac{\langle \zeta^n \rangle _c}{\langle \zeta^2 \rangle^{n/2}}.
\end{equation}
For the first non-trivial reduced cumulants, skewness and kurtosis, one has,
respectively,
\begin{equation}
D_3 = \frac{8 A^3\langle \chi^2 \rangle^3}{\left[ 2 A^2 \langle \chi^2 \rangle^2 \right]^{3/2}} = \sqrt{8},
\end{equation}
\begin{equation}
D_4 = \frac{48 A^4\langle \chi^2 \rangle^4}{\left[ 2 A^2 \langle \chi^2 \rangle^2 \right]^{4/2}} = 12.
\end{equation}
The general formula for $D_n$ is remarkably simple,
\begin{equation}
D_n = 2^{\frac{n}{2}-1} (n-1)!
\end{equation}

To find the PDF of the smoothed curvature fluctuations one must use the smoothed $\zeta$ field,
\begin{eqnarray}
\zeta_R({\bf x}) = A \int d^3 y W(|{\bf x - y}|/R) \chi^2({\bf y}) - \nonumber \\
- A \langle \chi^2 \rangle \int d^3 y W(|{\bf x - y}|/R).
\end{eqnarray}
Here, $W(x/R)$ is the window function. In the present paper we use the Gaussian window
function, defined by the equations
\begin{eqnarray}
W(x/R) = \frac{1}{V} e^{-\frac{x^2}{2 R^2}}, \quad V = (2\pi)^{3/2} R^3.
\end{eqnarray}

The general expressions for the cumulants of the PDF of the smoothed $\zeta$ field have
been derived in \cite{Matarrese:2000iz} using the path integral formalism. In this
formalism, authors of \cite{Matarrese:2000iz} expressed cumulants through the integrals
in $k$-space,
\begin{widetext}
\begin{eqnarray} \label{A20}
\langle \zeta_R^4 \rangle_c = 2^{n-1}(n-1)! A^n \int \frac{d^3 k_1}{(2\pi)^3} ...
\int \frac{d^3 k_n}{(2\pi)^3}
P_\chi(k_1) ... P_\chi(k_n) \times \qquad \qquad \qquad \qquad \qquad \qquad \nonumber \\ \times
\tilde W (|{\bf k_1 - k_2}| R) ...
\tilde W (|{\bf k_{n-1} - k_n}| R) \tilde W (|{\bf k_n - k_1}| R). \qquad
\end{eqnarray}
\end{widetext}
Here, $\tilde W(kR)$ is the window function in $k$-space, $\tilde W(kR)=e^{-k^2R^2/2}$,
$P_\chi(k)$ is the power spectrum of the $\chi$ field,
\begin{eqnarray}
P_\chi(k) = \frac{2 \pi^2}{k^3} {\cal P}_\chi(k).
\end{eqnarray}
As one can see from Eq. (\ref{A20}), values of the cumulants depend on the
$k$-dependence of the power spectrum of the $\chi$ field and on the window size $R$.
To study qualitatively the $R$-dependence of the cumulants it is more
convenient to use the expressions for $\langle \zeta_R^n\rangle_c$ through the
integrals in real (configuration) space \cite{Peebles:1998ph}. The
corresponding expression is
\begin{widetext}
\begin{eqnarray} \label{A22}
\langle \zeta_R^n\rangle_c = \int W(|{\bf x - x_1}|/R) W(|{\bf x - x_2}|/R)... W(|{\bf x - x_n}|/R)
\langle \zeta({\bf x_1}) \zeta({\bf x_2})...\zeta({\bf x_n}) \rangle_c d^3 x_1 d^3 x_2 ... d^3 x_n. \qquad
\end{eqnarray}
\end{widetext}
Here, the connected $n$-point function in real space is given by the product of
two-point correlation functions of $\chi$ field,
\begin{eqnarray} \label{A23}
\langle \zeta({\bf x_1}) ...\zeta({\bf x_n}) \rangle_c \sim \xi_\chi(x_{12})... \xi_\chi(x_{n1}),
\end{eqnarray}
\begin{eqnarray} \label{A24}
\xi_\chi(x_{ij}) = \int \limits_{0}^{\infty} {\cal P}_\chi(k) \frac{\sin (k x_{ij})}{k x_{ij}} \frac{dk}{k}.
\end{eqnarray}
In Eqs. (\ref{A23}, \ref{A24}) we use the notation $x_{ij}=|{\bf x}_i - {\bf x}_j|$.

If the power spectrum of the $\chi$ field has a form
\begin{equation}
{\cal P}_\chi\sim k^{t_\chi}, \;  t_\chi > 0,
\end{equation}
it follows from Eq. (\ref{A24}) that $\xi_\chi(x_{ij}) \sim x_{ij}^{-t_\chi}$, and
\begin{equation}
\langle \zeta({\bf x_1}) ...\zeta({\bf x_n}) \rangle_c \sim (x_{12} x_{23}\; ...\; x_{n1})^{-t_\chi}.
\end{equation}
Integrals in Eq. (\ref{A22}) converge, if $0 < t_\chi < 2.5$, and scale with the window size $R$.
Therefore, there is the proportionality $\langle \zeta_R^n\rangle_c \sim R^{-n t_\chi}$,
and, as a result, the reduced cumulants almost don't depend on the smoothing scale \cite{Peebles:1998ph},
\begin{equation} \label{A27}
D_{n,R} = \frac{\langle \zeta_R^n \rangle_c}{\langle \zeta_R^2\rangle^{n/2}} \sim
\frac{R^{-n t_\chi}}{R^{-2 t_\chi \frac{n}{2}}} \sim R^0.
\end{equation}
The weak dependence of the reduced cumulants on $R$ suggests that the PDF of the smoothed $\zeta$
field can be written in the form analogous to Eq. (\ref{A9}) \cite{Peebles:1998ph},
\begin{equation}
p_{\zeta,R}(\zeta_R) = \frac{1}{\langle \zeta_R^2 \rangle^{1/2}}
 \tilde p\left(\frac{\zeta_R}{\langle \zeta_R^2 \rangle^{1/2}}\right) \equiv
 \frac{1}{\langle \zeta_R^2 \rangle^{1/2}} \tilde p(\tilde \nu),
\end{equation}
$\tilde \nu \equiv \zeta_R/\langle \zeta_R^2 \rangle^{1/2}$.
Indeed, the reduced central moments for this PDF, which are given by the relation
\begin{eqnarray} \label{A29}
\frac{\langle \zeta_R^n \rangle}{\langle \zeta_R^2 \rangle^{n/2}} =
 \int \frac{\zeta_R^n}{\langle \zeta_R^2 \rangle^{n/2}}
  \frac{1}{\langle \zeta_R^2 \rangle^{1/2}}\tilde p
  \left( \frac{\zeta_R}{ \langle \zeta_R^2 \rangle^{1/2} } \right) d\zeta_R = \nonumber \\ =
 \int \nu_R^n \tilde p (\nu_R) d \nu_R, \qquad \qquad
\end{eqnarray}
have a form which is independent on the smoothing scale, in accordance with Eq. (\ref{A27}).

Evidently, the reduced cumulants which are connected with the reduced central moments
by a relation analogous to (\ref{A13}) also have this property.

Quantitative values of $D_{n,R}$ are different for different values of the power
spectrum index $t_\chi$ (even if $D_{n,R}$ almost do not depend on $R$). One can
expect, however, that if the positive tilt of the $\chi$-spectrum is not too
large, $t_\chi \lesssim 1$, the approximate equality
\begin{equation} \label{A30}
D_{n,R} \approx D_n
\end{equation}
takes place. This problem had been studied, for the case $n=3$, in \cite{Seto:2001mg},
and, for the case $n=4$, in \cite{White:1998da}. It had been shown in
\cite{Seto:2001mg, White:1998da} that, really, if $t_\chi$ is not small enough
(e.g., if $t_\chi=2$) the cumulants $D_{n,R}$ are comparatively small,
$D_{n,R} \ll D_n$, but they are close to $D_n$ in the limit $t_\chi \lesssim 1$
(just this limit is of interest for us in the present work).

Assuming that Eq. (\ref{A30}) holds for all $n$ (i.e., that the reduced cumulants are the same
in cases with smoothing and without smoothing), one can use for the PDF of the smoothed
$\zeta$ field the expression
\begin{equation}
p_{\zeta, R}(\zeta_R) = \frac{1}{\langle \zeta_R^2 \rangle^{1/2}} p(\tilde \nu),
\end{equation}
where $p(\tilde \nu)$ is given by Eq. (\ref{A10}), with a substitution $\nu \to \tilde\nu$. In
this approximation, the effects of the smoothing come only through the
variance $\langle \zeta_R^2 \rangle^{1/2}$ while the shape of the PDF is the
same as in the non-smoothing case.

The variance, $\langle \zeta_R^2 \rangle^{1/2} \equiv \sigma_\zeta(R)$, is given by
the expression followed from the general formula (\ref{A22}):
\begin{equation} \label{A32}
\langle \zeta_R^2 \rangle = \frac{2 A^2}{(2\pi)^6} \int dk dk' P_\chi(k) P_\chi(k')
\tilde W(|{\bf k} - {\bf k'}| R)^2.
\end{equation}

Note, for completeness, that moments of the PDF of the $\zeta$ field are simply connected
with {\it polyspectra} of the $\zeta$ field. In particular, using the definition
\begin{equation}
\langle \zeta({\bf k_1}) \zeta({\bf k_2}) \rangle = (2\pi)^3 \delta_D({\bf k_1}+{\bf k_2}) P_\zeta(k_1),
\end{equation}
one can obtain from Eq. (\ref{A32}) the simple formula for the variance:
\begin{equation} \label{A34}
\langle \zeta_R^2 \rangle = \sigma_\zeta^2(R) = \int \limits_0^{\infty}
 \tilde W^2(kR) {\cal P}_\zeta(k) \frac{dk}{k}.
\end{equation}


\end{document}